\documentclass[]{aa}

\usepackage{color}
\usepackage{txfonts}
\usepackage{graphicx} 
\usepackage{natbib} 
\usepackage{longtable} 
\usepackage{amssymb}
\usepackage{lscape}
\usepackage{rotating}
\usepackage[english]{babel}
\bibpunct{(}{)}{;}{a}{}{,}

\begin{document}

\title{The white dwarf population of NGC~6397}

\author{Santiago Torres\inst{1,2}\and
        Enrique Garc\'\i a--Berro\inst{1,2}\and 
        Leandro G. Althaus\inst{3,4}\and
        Mar\'\i a E. Camisassa\inst{3,4}}
        
\institute{Departament de F\'\i sica Aplicada, 
           Universitat Polit\`ecnica de Catalunya, 
           c/Esteve Terrades 5, 
           08860 Castelldefels, 
           Spain
           \and
           Institute for Space Studies of Catalonia, 
           c/Gran Capita 2--4, 
           Edif. Nexus 104, 
           08034 Barcelona, 
           Spain
           \and
           Facultad de Ciencias Astron\'omicas y Geof\'isicas, 
           Universidad Nacional de La Plata,
           Paseo del Bosque s/n, 
           1900 La Plata, 
           Argentina
           \and
           Instituto de Astrof\'isica de La Plata, UNLP-CONICET,
           Paseo del Bosque s/n, 
           1900 La Plata, 
           Argentina}

\date{\today}

\titlerunning{The white dwarf population of NGC~6397}
\authorrunning{Torres et al.}

\offprints{E. Garc\'\i a--Berro  }


\abstract 
          {NGC~6397 is one of the  most interesting, well observed and
            theoretically  studied  globular clusters.   The  existing
            wealth of observations allows  us to study the reliability
            of the  theoretical white  dwarf cooling sequences  of low
            metallicity  progenitors, to  determine  its  age and  the
            percentage  of unresolved  binaries, and  to assess  other
            important characteristics  of the cluster, like  the slope
            of the  initial mass  function, or  the fraction  of white
            dwarfs with hydrogen deficient atmospheres.}
          {We present a population synthesis  study of the white dwarf
            population of NGC~6397.  In particular, we study the shape
            of  the  color-magnitude  diagram, and  the  corresponding
            magnitude and color distributions.}
          {We  do  this using  an  up-to-date  Monte Carlo  code  that
            incorporates   the  most   recent  and   reliable  cooling
            sequences and  an accurate  modeling of  the observational
            biases.}
          {We find a good agreement between our theoretical models and
            the  observed  data.  In  particular,  we  find that  this
            agreement is  best for  those cooling sequences  that take
            into account  residual hydrogen  burning. This  result has
            important  consequences for  the  evolution of  progenitor
            stars during the thermally-pulsing asymptotic giant branch
            phase, since  it implies that appreciable  third dredge-up
            in low-mass,  low-metallicity progenitors is  not expected
            to occur.  Using a standard  burst duration of 1.0~Gyr, we
            obtain    that    the    age    of    the    cluster    is
            $12.8^{+0.50}_{-0.75}$~Gyr.    Larger    ages   are   also
            compatible with  the observed  data, but  then unrealistic
            longer durations  of the  initial burst of  star formation
            are needed to fit the luminosity function.}
          {We  conclude that  a correct  modeling of  the white  dwarf
            population of globular clusters,  used in combination with
            the  number  counts of  main  sequence  stars provides  an
            unique tool to model the properties of globular clusters.}


\maketitle


\section{Introduction}

White dwarfs are the remnants of  the evolution of single stars of low
and intermediate mass. Consequently, they gather important information
not only  about the evolution of  their progenitor stars, but  also on
the properties of their parent populations. Moreover, since in general
terms,  the  evolution  of  white   dwarf  stars  is  relatively  well
understood, the ensemble  properties of their populations  can be used
to test  the validity of physical  theories that cannot be  done using
terrestrial facilities \citep{varG1, varG2, axions1, axions0, axions2,
axions3}.   This has  been possible  because  we now  have a  profound
knowledge  of   the  physics  governing  white   dwarf  interiors  and
envelopes, which has resulted in accurate cooling tracks, as well as a
wealth of observational data.  Even more, the quality of the available
observational  data  has allowed  us  to  check  the accuracy  of  our
description of the main physical processes involved in their evolution
-- see, for instance \citet{Althaus2010} for a recent review.  In this
vein, it is important to realize that due to the conceptual simplicity
of the white  dwarf populations of globular  clusters their properties
can  be  used  to  test  white dwarf  evolutionary  models.   This  is
particularly  interesting because  it  has been  recently shown  that,
under  controlled conditions,  theoretical white  dwarf cooling  times
obtained using  very different numerical  codes and techniques  do not
differ by  much. Hence,  white dwarf evolutionary  times should  be at
least as reliable as main sequence lifetimes \citep{Salaris2013}.

Several studies have taken advantage of all these properties, and have
determined  accurate ages  of both  open and  globular clusters.   For
instance,  \cite{Gar_etal_2010}   determined  the  age  of   the  old,
metal-rich,   well-populated,   open   cluster  NGC~6791   using   the
termination   of  the   cooling  sequence   \citep{Bedin05,  Bedin08a,
  Bedin08b}, and solving the  long-standing problem of the discrepancy
between the main  sequence and white dwarf ages.  But  this is not the
only cluster  for which  we have good  age determinations  using white
dwarf  evolutionary  sequences.  Other  examples  are  the young  open
clusters   M~67  \citep{Bellini},   NGC~2158  \citep{Bedin2010},   and
NGC~6819     \citep{Bedin2015},    the     globular    clusters     M4
\citep{Hansen_2002, Bedin2009}, $\omega$ Centauri \citep{Bellini2013},
and NGC~6397  \citep{Hansen_2013}, or the metal-rich  globular cluster
47~Tuc \citep{Goldsbury_2012, Gar_etal_2014}.

NGC~6397 is  the second-nearest globular  cluster to the Sun,  and has
been thoroughly  observed by the  {\sl Hubble Space  Telescope}. Thus,
for it we have high-quality deep  images that have allowed us to study
not only  the lower main  sequence, but  also the white  dwarf cooling
sequence.   NGC~6397  is old  and  metal-poor,  being its  metallicity
[Fe/H]$=-1.8$ \citep{Hansen_2013},  although in the  recent literature
there   exists    some   discrepancy    about   its    precise   value
\citep{Richer2008}. The same happens for  the age of the cluster.  For
instance, \cite{Hansen_2007}  analyzed the position of  the cut-off of
the white  dwarf luminosity  function and, comparing  with theoretical
cooling   models,  derived   an   age  $T_{\rm   c}=11.47\pm0.47$~Gyr.
Similarly,     \cite{Winget_2009},    fitted     simultaneously    the
main-sequence, the pre-white dwarf and  the white dwarf regions of the
color-magnitude    diagram,    and    obtained    an    age    $T_{\rm
c}=12.0^{+0.5}_{-1.0}$~Gyr.   These ages,  which  are  based on  white
dwarf evolutionary  models, need  to be  compared with  those obtained
fitting the  luminosity of  the main-sequence  turn-off (MSTO)  -- see
\cite{Richer2008}  for  a  discussion  of different  methods  and  age
estimates of NGC~6397.   In particular, it is  worth highlighting that
using   this   method   \cite{Gratton_2003}    derived   an   age   of
$13.9\pm1.1$~Gyr.  This  prompted \cite{Strickler_2009} to  claim that
this age determination was compatible with the possible existence of a
putative  population of  very old  helium white  dwarfs. On  the other
hand,     \cite{Twarog2000}      obtained     $12.0\pm0.8$~Gyr     and
\cite{Chaboyer_2001} derived  $13.4\pm0.8~$Gyr. Thus, the  precise age
of NGC~6397 remains  unclear, and needs to  be independently evaluated
using  the  most  recent  white   dwarf  evolutionary  tracks  of  the
appropriate metallicity.   To this  we add  that NGC~6397  shares some
other  interesting   characteristics  with  other   Galactic  globular
clusters, such  as the  unusual lithium enhancement  of some  of their
stars --  see \cite{Pasquini_2014}  and references  therein --  or the
possible   existence   of    multiple   populations   \citep{Cris2010,
Milone2012}.  Thus,  deriving an  independent white dwarf  cooling age
for this globular cluster is of crucial importance.

Moreover, NGC~6397  is one of the  few globular cluster for  which, in
addition  to its  color-magnitude diagram,  we also  have an  accurate
white dwarf  luminosity function, thus  allowing us to  investigate in
detail    those    issues   previously    mentioned.     Specifically,
\cite{Winget_2009}  used  the  shape  of the  white  dwarf  luminosity
function of NGC~6397 to constrain  important properties of white dwarf
interiors.  They found  that to account for the observed  shape of the
white dwarf luminosity  function of NGC~6397 either a  low oxygen mass
fraction in the  inner carbon-oxygen core of typical  white dwarfs was
needed, or the crystallization  temperature of the carbon-oxygen dense
binary  plasma should  be  significantly  larger than  that  of a  one
component plasma.  The reason for  this is that when the carbon-oxygen
plasma  crystallizes  the  oxygen  abundance in  the  solid  phase  is
enhanced  \citep{GB88b,   GB88a,  Horowitz2010}.   This  leads   to  a
redistribution of oxygen  in the core of the white  dwarf, because the
inner  regions are  enriched in  oxygen, while  the outer  ones become
oxygen-poor.   Since  oxygen  is  slightly heavier  than  carbon  this
chemical separation process  releases gravitational energy, increasing
the  cooling times  \citep{Isern1997, Isern2000}.   \cite{Winget_2009}
found that  to fit  the observations the  effects of  phase separation
upon  crystallization   should  be   minimized.   Hence,   either  the
gravitational energy  should be  released at larger  luminosities (or,
equivalently, core temperatures), or the  oxygen abundance in the deep
interior of typical white dwarfs should  be smaller. This, in turn, is
an important topic, as a low  oxygen mass fraction could be indicative
of    a    small    cross    section   of    the    poorly    measured
C$^{12}(\alpha,\gamma)$O$^{16}$   nuclear  reaction,   since  it   has
implications   for   the   evolution  of   white   dwarf   progenitors
\citep{Salaris1997}.   Moreover,  since  the   precise  value  of  the
C$^{12}(\alpha,\gamma)$O$^{16}$ cross section is  still the subject of
an  active   debate  --  see,  for   instance,  \cite{Avila2015},  and
references  therein --  any additional  piece of  evidence helping  in
constraining it is valuable.

Furthermore, recent theoretical models  predict that at moderately low
luminosities a  phase of  stable hydrogen nuclear  burning in  a shell
ensues  in  the  atmospheres  of  white  dwarfs  descending  from  low
metallicity  progenitors   \citep{Miller_etal_2013}.   More  recently,
\cite{camissasa}     have     expanded     the     calculations     of
\cite{Miller_etal_2013}   and   have   shown   that   for   progenitor
metallicities between 0.00003 and 0.001,  and in the absence of carbon
enrichment due  to the  occurrence of a  third dredge-up  episode, the
resulting hydrogen envelope  of low-mass white dwarfs  is thick enough
to make stable hydrogen burning  the most important energy source even
at low luminosities. Although in some cases this may not be a relevant
source of  energy, it may  have non-negligible effects  when computing
the ages of very old clusters of low metallicity.

In  this paper  we  analyze the  white dwarf  population  of the  old,
metal-poor globular cluster NGC~6397.  We use an up-to-date population
synthesis code based on Monte  Carlo techniques, that incorporates the
most  recent  and  reliable   cooling  sequences,  a  state-of-the-art
description of the main properties of NGC~6397, as well as an accurate
modeling  of the  observational  biases.  Our  paper  is organized  as
follows.  In Sects.~\ref{sec:popsyn} and \ref{sec:obsdat} we describe,
respectively,  our   population  synthesis   code  and   the  observed
sample. Sect.~\ref{sec:res} is  devoted to analyze the  results of our
simulations.   Specifically, we  study the  role of  residual hydrogen
burning,  the  effects  of  the  slope of  the  adopted  initial  mass
function, the possible effects of  mass segregation on the white dwarf
population,  and  the  fraction  of white  dwarf  with  hydrogen  poor
atmospheres.  All  this is done  by performing a  detailed statistical
analysis of the luminosity function,  of the color distribution and of
the color-magnitude diagram of the white dwarf population of NGC~6397.
Finally, in  Sect.~\ref{sec:con} our  main results are  summarized and
our conclusions are drawn.

\section{The population synthesis code}
\label{sec:popsyn}

Our synthetic population code is  based on Monte Carlo techniques, and
has  been  extensively  used  for studying  the  disk  \citep{Gar1999,
Tor2001} and halo \citep{Tor2002, Gar2004} populations of single white
dwarfs.  An  improved version  of this  population synthesis  code has
been recently employed to model  the properties of the disk population
of    white     dwarf    plus     main    sequence     binary    stars
\citep{Cam2014}. Finally, and  most relevant for this  study, our code
has also been successfully used  to model Galactic open clusters, such
as   NGC~6791  \citep{Gar_etal_2010,   Gar_etal_2011},  and   globular
clusters, like 47~Tuc  \citep{Gar_etal_2014}. Detailed descriptions of
our Monte Carlo simulator can be  found in these papers. Thus, in this
section we only summarize its most  salient features, and we refer the
interested reader for more details to the previously mentioned works.

Synthetic  main  sequence stars  are  randomly  drawn according  to  a
Salpeter-like  initial  mass  function.   In this  work  we  used  the
so-called ``universal'' mass function  of \cite{Kroupa_2001}.  For the
mass range relevant to our study this initial mass function is totally
equivalent to  a two-branch  power law  with exponent  $-\alpha$, with
$\alpha = 1.3$ for $0.08 \leq  M/M_{\sun}<0.5$ and $\alpha = 2.30$ for
$M/M_{\sun}  \geq 0.5$.   However, for  the sake  of completeness  and
simplicity we  also use  a classical  Salpeter-like \citep{Salpeter55}
initial mass function in Sect.~\ref{subsec:imf}. This mass function is
a power law with just one index, which we consider a free parameter to
fit the  observations.  The selected range  of masses at the  zero age
main sequence is that necessary  to produce white dwarf progenitors of
suitable masses  for NGC~6397.  In particular,  a lower limit of  $M >
0.5\, M_{\sun}$ guarantees that enough white dwarfs are produced for a
broad range of cluster ages. Within  this mass range both initial mass
functions  are   totally  equivalent   when  the  standard   value  of
$\alpha=2.35$ is adopted.

\begin{table*}[t]
\begin{center}
\caption{$\chi^2$ test of the  luminosity function, color distribution
  and  color-magnitude diagram  for  different ages,  duration of  the
  burst of star  formation and binary fractions.  We  list the reduced
  and   normalized  value   of  $\chi^2$,   that  is   the  value   of
  $\chi^2_{\nu}$ over its minimum value.}
\label{t:totalchic}
\tiny
\begin{tabular}{lccccccccc}
\hline
\hline
 $\chi^2$ test & \multicolumn{3}{c}{ $\chi^2_{\rm F814W,\nu}/\chi^2_{\rm min}$ } &  \multicolumn{3}{c}{ $\chi^2_{\rm F606W-F814W,\nu}/\chi^2_{\rm min}$} & \multicolumn{3}{c}{ $\chi^2_{\rm N,\nu}/\chi^2_{\rm min}$}  \\
\hline
&\multicolumn{9}{c}{$T{\rm c}=11.0\,$(Gyr)}\\
\cline{2-10}
$\Delta t\,$(Gyr) & 1.0 & 2.0 & 3.0 & 1.0 & 2.0 & 3.0 & 1.0 & 2.0 & 3.0 \\
$f_{\rm BIN}=0.00$ &  9.42  & 11.31  & 12.68  &  2.69  &  3.60  &  4.67  &  4.17  &  5.33  &  5.88  \\
$f_{\rm BIN}=0.02$ &  9.72  & 11.48  & 12.79  &  2.62  &  3.54  &  4.47  &  4.18  &  5.29  &  5.90  \\
$f_{\rm BIN}=0.04$ &  9.71  & 11.38  & 12.86  &  2.65  &  3.56  &  4.63  &  4.16  &  5.31  &  5.97  \\
$f_{\rm BIN}=0.06$ &  9.62  & 11.53  & 12.84  &  2.58  &  3.54  &  4.49  &  4.16  &  5.36  &  5.82  \\
\hline
&\multicolumn{9}{c}{$T{\rm c}=12.0\,$(Gyr)}\\
\cline{2-10}
$\Delta t\,$(Gyr) & 1.0 & 2.0 & 3.0 & 1.0 & 2.0 & 3.0 & 1.0 & 2.0 & 3.0 \\
$f_{\rm BIN}=0.00$ &  4.43  &  5.92  &  7.33  &  1.24  &  1.68  &  2.40  &  1.71  &  2.50  &  3.28  \\
$f_{\rm BIN}=0.02$ &  4.49  &  5.95  &  7.54  &  1.28  &  1.62  &  2.37  &  1.78  &  2.44  &  3.23  \\
$f_{\rm BIN}=0.04$ &  4.44  &  6.02  &  7.52  &  1.26  &  1.67  &  2.36  &  1.72  &  2.41  &  3.20  \\
$f_{\rm BIN}=0.06$ &  4.38  &  6.07  &  7.39  &  1.30  &  1.63  &  2.31  &  1.71  &  2.45  &  3.13  \\
\hline
&\multicolumn{9}{c}{$T{\rm c}=13.0\,$(Gyr)}\\
\cline{2-10}
$\Delta t\,$(Gyr) & 1.0 & 2.0 & 3.0 & 1.0 & 2.0 & 3.0 & 1.0 & 2.0 & 3.0 \\
$f_{\rm BIN}=0.00$ & 1.95  &  2.11  &  2.86  &  1.24  &  1.08  &  1.16  &  1.43  &  1.17  &  1.48  \\
$f_{\rm BIN}=0.02$ & 2.05  &  2.14  &  2.92  &  1.34  &  1.07  &  1.16  &  1.42  &  1.10  &  1.45  \\
$f_{\rm BIN}=0.04$ & 1.90  &  2.09  &  2.94  &  1.32  &  1.08  &  1.15  &  1.38  &  1.12  &  1.41  \\
$f_{\rm BIN}=0.06$ & 1.87  &  2.07  &  2.97  &  1.34  &  1.11  &  1.17  &  1.37  &  1.13  &  1.40  \\
 
\hline
&\multicolumn{9}{c}{$T{\rm c}=14.0\,$(Gyr)}\\
\cline{2-10}
$\Delta t\,$(Gyr) & 1.0 & 2.0 & 3.0 & 1.0 & 2.0 & 3.0 & 1.0 & 2.0 & 3.0 \\
$f_{\rm BIN}=0.00$ & 4.00  &  2.04  &  1.18  &  2.40  &  1.75  &  1.29  &  2.59  &  1.93  &  1.34  \\
$f_{\rm BIN}=0.02$ & 3.81  &  2.10  &  1.11  &  2.52  &  1.82  &  1.28  &  2.60  &  1.84  &  1.39  \\
$f_{\rm BIN}=0.04$ & 3.72  &  1.90  &  1.08  &  2.46  &  1.78  &  1.30  &  2.61  &  1.81  &  1.28  \\
$f_{\rm BIN}=0.06$ & 3.61  &  1.82  &  1.07  &  2.52  &  1.79  &  1.33  &  2.54  &  1.86  &  1.38  \\
\hline
\end{tabular}
\end{center}
\end{table*}

We adopt  an age of  the cluster, $T_{\rm  c}$, which falls  within an
interval going from 11.0 to 14.5~Gyr.  This is done to account for the
widest relevant range of age estimates  of NGC~6397.  We also employ a
star formation rate consisting of a burst of duration $\Delta t$.  The
width of the  burst covers a sufficiently large range  of values, from
0.0  to  4.0~Gyr,  to  ensure that  all  possibilities  are  explored.
Finally, we adopted a maximum fraction of unresolved double-degenerate
binaries $f_{\rm  BIN}\le 8\%$.  This  upper limit is  consistent with
the    expected   values    reported    by   \cite{Hansen_2007}    and
\cite{Davis_2008}.   The masses  of the  individual components  of the
binary   system   were   randomly    drawn   according   to   a   flat
distribution. However, the results are  not sensitive to the choice of
the mass distribution.   We also considered that half  of these binary
systems are composed by a  helium-core white dwarf and a carbon-oxygen
one,  whereas for  the  rest of  the systems  the  two components  are
typical carbon-oxygen white dwarfs. For white dwarfs with helium cores
we adopted a mean mass of $0.4\,M_{\sun}$ with a Gaussian deviation of
$0.15\,M_{\sun}$

\begin{figure}[t]
   \resizebox{1.1\hsize}{!}
   {\includegraphics[width=\columnwidth]{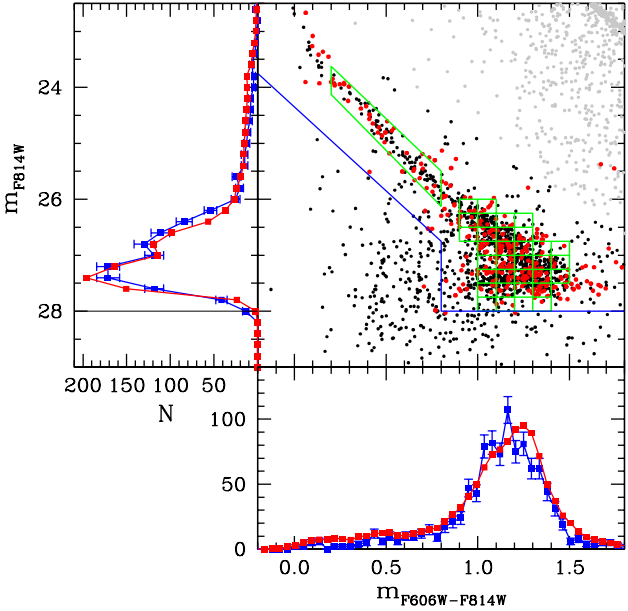}}
   \caption{White dwarf  luminosity function,  color-magnitude diagram
     and color distribution of NGC~6397,  for our best-fit model. Grey
     dots  represent   observed  main   sequence  stars,   black  dots
     correspond to white  dwarfs, while red points  denote the results
     of our simulations.   The green squares represent  the regions of
     the  color-magnitude diagram  for  which we  peformed a  $\chi^2$
     test, while the blue thin lines correspond to the cuts adopted to
     compute  the  distributions. The  red  curves  correspond to  the
     simulated  distributions, while  the blue  ones are  the observed
     distributions computed using our cuts.  See the online edition of
     the journal for a color version of this figure, and the main text
     for additional details.}
\label{f:WDLF_CON}
\end{figure}

\begin{figure}[t]
\includegraphics[trim = 40mm 0mm 0mm 0mm, clip, width=1.5\columnwidth]{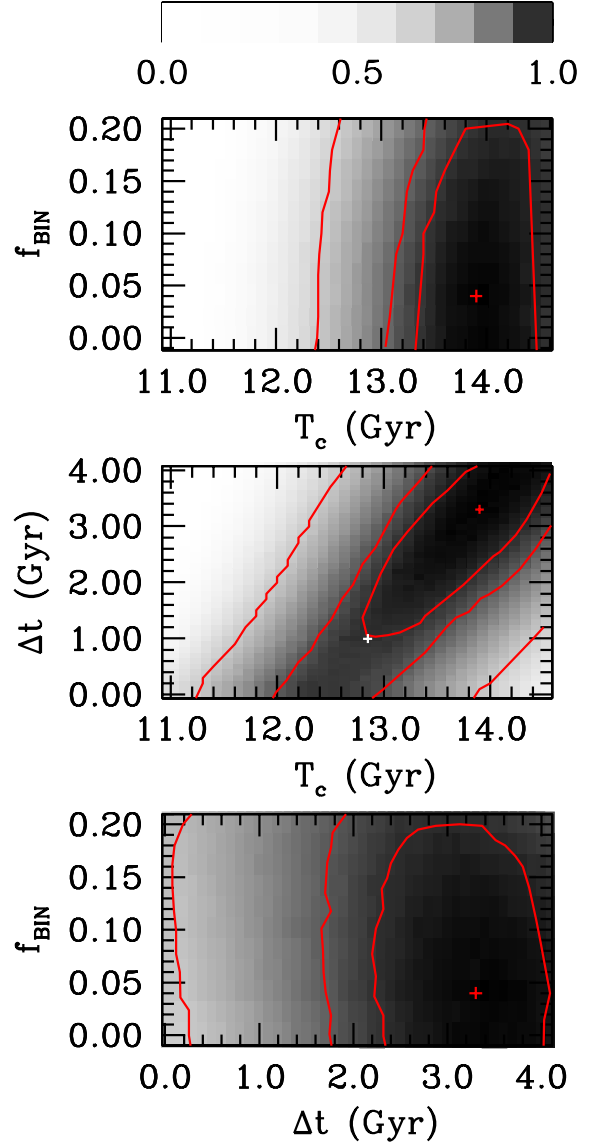}
   \caption{Probability  density distribution  for different  pairs of
     free  parameters  (gray scale).   Marked  with  a cross  are  the
     best-fit values,  while the red  lines represents the  68\%, 90\%
     and 95\%  confidence levels. The  white cross corresponds  to our
     reference model.}
\label{f:PDIS_CON}
\end{figure}

Once  we know  which  stars had  time  to evolve  to  white dwarfs  we
interpolate their  photometric properties  using a set  of theoretical
cooling sequences \citep{Miller_etal_2013, camissasa} for white dwarfs
with  hydrogen-rich  atmospheres,  in  accordance  with  the  observed
spectroscopic classification of the white  dwarfs of the cluster. This
set of evolutionary sequences consists of a grid of cooling tracks for
carbon-oxygen white  dwarfs derived from evolutionary  calculations of
progenitors with  metallicities ranging from $Z=1.0\times  10^{-4}$ to
$Z=5.0\times 10^{-4}$.  We note here that these evolutionary sequences
were  evolved  self-consistently  from  the zero  age  main  sequence,
through the giant phase, the thermally pulsing asymptotic giant branch
(AGB)  and  mass-loss  phases,  and  ultimately  to  the  white  dwarf
stage. We also emphasize that  these evolutionary sequences provide us
with a  self-consistent initial-to-final mass relationship.   For more
massive oxygen-neon white dwarfs we  employed the cooling sequences of
\cite{One2} and \cite{ONe1}.  For helium-core white dwarfs we used the
cooling  sequences of  \cite{Serenelli_2001}, while  for non-DA  white
dwarfs  we used  the cooling  tracks of  \cite{Bergeron2011}, and  the
colors of \cite{Sirianni}.   An important detail here is  that to keep
consistency, the masses  of the resulting synthetic  white dwarfs were
interpolated  using the  initial-to-final mass  relationships obtained
from the previously  mentioned set of full  evolutionary sequences for
metal poor  progenitors.  Nevertheless, for the  sake of completeness,
we   also  ran   some   simulations  for   which  the   semi-empirical
initial-to-final   mass    relationship   of   field    white   dwarfs
\citep{Catalan2008b, Catalan2008a} was used, obtaining essentially the
same results.

Photometric errors  were assigned  randomly according to  the observed
distribution.   Specifically,  for  each  synthetic  white  dwarf  the
photometric errors  are drawn within a  hyperbolically increasing band
limited  by  $\sigma_{\rm l}=0.2\left(m_{\rm  F814W}-31.0\right)^{-2}$
and  $\sigma_{\rm   u}-0.06=1.7\left(m_{\rm  F814W}-31.0\right)^{-2}$,
which fits well the observations of \cite{Hansen_2007} for the $m_{\rm
F814W}$ magnitude.  Similar expressions were  employed for the rest of
the magnitudes.

\section{The observed data set}
\label{sec:obsdat}

The observed data to which  we compare our population synthesis models
was  obtained from  \cite{Hansen_2007}, and  consists of  a series  of
images taken with the Advanced  Camera for Surveys (ACS) field located
$5^\prime$  southeast  of  the  cluster core.   Zero  points  for  the
instrumental magnitudes are 32.414 for  $m_{\rm F814W}$ and 33.321 for
$m_{\rm  F606W}$. Besides,  an absolute  sharp parameter  smaller than
$0.02$  and round  parameter smaller  than $0.02$  was used  to remove
extended   sources.   In   Fig.~\ref{f:WDLF_CON}  we   show  the   raw
color-magnitude  diagram of  the sample.   Those objects  with $m_{\rm
F814W}<2.3(m_{\rm  F606W}-m_{\rm  F814W})+22.5$   are  far  below  the
main-sequence and can  safely be considered as  candidate white dwarfs
(black  dots in  the top-right  panel of  Fig.~\ref{f:WDLF_CON}).  For
magnitude  27  the  completeness  of  the  sample  is  80\%,  and  for
magnitudes larger than this value steadily decreases, reaching 50\% at
magnitude 28.   The black line  in the  top-left panel of  this figure
represents this limit.

A third epoch set of observations  to get deeper proper motions, which
are  especially  useful for  discriminating  white  dwarfs with  faint
magnitudes, has not been published yet.  Consequently, and in order to
avoid  contamination from  galaxies,  we  apply a  set  of color  cuts
(represented  by the  blue  lines in  the  color-magnitude diagram  of
Fig.~\ref{f:WDLF_CON}).   Those   objects  below  the  blue   line  of
Fig.~\ref{f:WDLF_CON} are not considered  in our analysis. Finally, we
build  a  ``clean''  luminosity  function  and  a  color  distribution
function  --   left  and   bottom  panels   of  Fig.~\ref{f:WDLF_CON},
respectively -- taking into account  these additional color cuts (blue
lines).

\section{Results}
\label{sec:res}

\subsection{The color-magnitude diagram}
\label{subsec:onepop}

We first discuss the overall shape of the color-magnitude diagram, and
the corresponding  distributions of  magnitudes and colors.   All this
information is displayed in Fig.~\ref{f:WDLF_CON} for a model in which
we  adopt  an  age  $T_{\rm c}=12.8$~Gyr,  a  burst  duration  $\Delta
t=1.0$~Gyr, and  a fraction  of binaries  $f_{\rm BIN}=4\%$,  which we
consider  our reference  model.  The  top-right panel  of this  figure
shows the observed stars, as well  as the synthetic white dwarfs.  The
results of our synthetic population calculations are represented using
red  dots in  the color-magnitude  diagram, whereas  the corresponding
magnitude  and  color   distributions  are  shown  as   red  lines  in
Fig.~\ref{f:WDLF_CON}.  To obtain the color-magnitude diagram shown in
this  figure we  fitted  the  position of  the  bright  branch of  the
degenerate cooling  sequence, obtaining  an apparent  distance modulus
$(m-M)_{\rm F814W}=12.42^{+0.05}_{-0.09}$  and a color  excess $E({\rm
F606W-F814W})=0.22\pm0.02$.    These  values   agree  with   those  of
\cite{Richer2013}, and references therein. 

As can  be seen,  the agreement  between our  reference model  and the
observations is excellent.  However, in our modeling we  went one step
beyond and performed a $\chi^2$ analysis using the following strategy.
We computed independent $\chi^2$ tests for the magnitude ($\chi^2_{\rm
F814W}$) and color ($\chi^2_{{\rm F606W}-{\rm F814W}}$) distributions.
Additionally, we calculated the number  of white dwarfs inside each of
the    green    boxes    in    the    color-magnitude    diagram    of
Fig.~\ref{f:WDLF_CON} --  which are the  same regions of  this diagram
used by \cite{Hansen_2013} to  compare observations and simulations --
and we performed an additional  $\chi^2$ test, $\chi^2_{N}$.  Then, we
obtained for each of the  three independent $\chi^2$ tests the reduced
value  of $\chi^2$,  $\chi^2_{\nu}=\chi^2/{\nu}$, where  $\nu$ is  the
number of  degrees of  freedom, $\nu=N-p-1$, being  $N$ the  number of
bins and $p$ the number of free parameters.  Then the reduced $\chi^2$
values were normalized to the minimum  value for each of the tests and
added in quadrature.  Employing this procedure only one estimator must
be  minimized to  find the  model  that best  fits all  distributions.
Clearly, this strategy  is equivalent to a  maximum likelihood method,
hence the  results of both  approaches are consistently  similar.  For
illustrative   purposes,   in   Table~\ref{t:totalchic}  we   list   a
representative  set  of  values  for  the  different  tests  and  free
parameters studied here.  It is worth mentioning here that although we
have used a $\chi^2$ test, our final  aim is to estimate the values of
the  free parameters  that best  fit  the observed  data, rather  than
obtaining an absolute probability of  agreement of our models with the
observed data.

In Fig.~\ref{f:PDIS_CON}  we show using  a gray scale  the probability
distribution for different pairs of the studied free parameters -- the
cluster  age, $T_{\rm  c}$, the  burst duration,  $\Delta t$,  and the
binary fraction, $f_{\rm  BIN}$. Also shown, using red  lines, are the
curves  enclosing  the  regions  of 68\%,  90\%  and  95\%  confidence
level. Finally, our best fit model is  marked with a red cross and our
reference  model with  a white  one.  Using  the procedure  previously
detailed we found that the model  that best fits the observed data has
the  following  properties.   The  age  of  the   cluster  is  $T_{\rm
c}=13.9$~Gyr, the duration  of the burst of star  formation is $\Delta
t=3.3$~Gyr  and  the  binary  fraction   of  the  cluster  is  $f_{\rm
BIN}=4\%$. One  point of concern  regarding our best-fit model  is its
very  large age  spread, 3.3~Gyr.   This is  unusual in  most globular
cluster but  on two, $\omega$~Centauri  -- see \cite{Villanova}  for a
recent  discussion about  this  issue --  and M54  \citep{Siegel2007}.
Nevertheless, we remark that the previous theoretical efforts to model
the white  dwarf luminosity  function of  NGC~6397 could  only achieve
satisfactory  fits when  a  constant star  formation  rate during  the
entire life of the cluster was employed \citep{Winget_2009}.  However,
we  judge that  the burst  duration obtained  using this  procedure is
totally  unrealistic,  and that  our  reference  model, for  which  we
adopted a  typical burst  duration of 1.0~Gyr  is more  realistic.  We
elaborate on the reasons for this below.

Inspection  of  the top  and  bottom  panels of  Fig.~\ref{f:PDIS_CON}
reveals that there  is no correlation between the  adopted fraction of
binaries  and  the   duration  of  the  burst  or   the  cluster  age.
Accordingly, $f_{\rm  BIN}$ can be  considered as an  independent free
parameter.  Thus, its value has no major effects on the final results.
In contrast, the middle panel of Fig.~\ref{f:PDIS_CON} unveils a clear
unphysical correlation between  the duration of the burst  and the age
of  the cluster.   Actually, ages  shorter than  that of  our best-fit
model are possible provided that  smaller values of the burst duration
are adopted, and the reverse is also  true.  In fact, ages as short as
12.8~Gyr  are  compatible, at  the  95\%  confidence level,  with  the
observations  if  the  initial  burst of  star  formation  lasted  for
1.0~Gyr.  Even  a cluster age  of 12.4~Gyr  is compatible at  the 90\%
confidence level with  the observed data if the duration  of the burst
of  star formation  is 0.1~Gyr.   Moreover, if  $\Delta t=0.0$~Gyr  is
adopted    the   age    of    the   cluster    turns    out   to    be
$12.34^{+0.5}_{-0.4}$~Gyr,  in  agreement  with other  ages  estimates
\citep{Hansen_2013}. All this  is due to the very flat  maximum of the
probability distributions displayed in Fig.~\ref{f:PDIS_CON}.

Since this  is an important issue  we explored which is  the origin of
this unphysical  correlation between  the age of  the cluster  and the
duration of the episode of star formation computed using the available
white dwarf  luminosity function.  We  found that the main  reason for
this is that the observed stars  with $m_{\rm F814W}\ge 26$ have large
photometric errors. Thus,  the precise position of the  maximum of the
observed  white dwarf  luminosity function  cannot be  determined with
good  accuracy.  Hence,  age  determinations relying  on the  observed
luminosity   function   are    heavily   influenced   by   photometric
errors. Additionally, such an extended episode of star formation would
influence  the  morphology  of  the color-magnitude  diagram  of  main
sequence  stars   of  NGC~6397.    Nowadays  with   the  high-accuracy
photometric   data  collected   with   the   Hubble  Space   Telescope
\citep{Milone2012} this age  spread should be detectable,  and this is
not the  case.  In conclusion,  we judge that the  adopted photometric
dataset is  not yet appropriate  for this  kind of analysis,  and that
there are  solid reasons to suspect  that the duration of  the initial
burst  of  star formation  is  shorter  than  that obtained  from  our
analysis. Consequently, from  now on we stick to  our reference model,
in which a burst duration of 1.0~Gyr is adopted.

Finally, we  checked the  sensitivity of our  results to  the distance
modulus and the  reddening.  Specifically, we varied  the reddening by
$\pm 0.02$, the nominal errors in its determination, and we found that
the derived  age for the cluster  when our reference model  is adopted
varies  by $\pm  0.02$~Gyr.  We  repeated the  same procedure  for the
value of the distance  modulus and we found that in  this case the age
of  the cluster  varies  by $\pm  0.3$~Gyr.   Furthermore, fixing  the
duration  of  the  initial  burst  of star  formation  we  obtained  a
conservative estimate  of the error  associated to the derived  age of
the  cluster, and  also  to  the fraction  of  binaries.  We  obtained
$\sigma_{T_{\rm    c}}=^{+0.50}_{-0.75}$~Gyr,   and    $\sigma_{f_{\rm
BIN}}=^{+0.14}_{-0.04}$.   As a  final  analysis we  also studied  the
robustness of our results to  the completeness of the observed sample.
In  particular,  we discarded  20\%  of  synthetic white  dwarfs  with
magnitudes smaller than 26, while for  magnitudes between 26 and 28 we
assumed  that  the completeness  decreases  linearly  until 50\%,  and
removed the  corresponding synthetic  white dwarfs.  We  then computed
the  white dwarf  luminosity function  and color  distribution of  the
resulting sample, and  we found that the results  were essentially the
same.  We thus  conclude that the age derived for  our reference model
is robust.

\subsection{The role of residual nuclear burning}
\label{subsec:resh}

\begin{figure}[t]
   \resizebox{1.1\hsize}{!}
   {\includegraphics[width=\columnwidth]{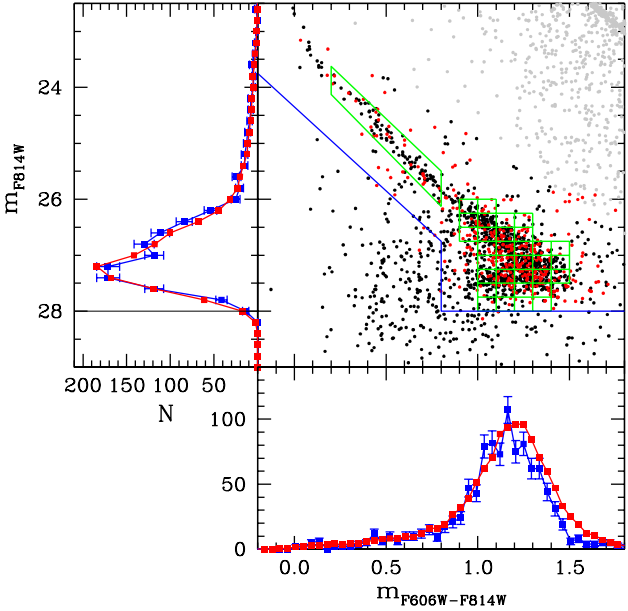}}
   \caption{Same as  Fig.~\ref{f:WDLF_CON} but  for the model  with no
     residual hydrogen  burning, see Sect.~\ref{subsec:resh}.}
\label{f:WDLF_SIN}
\end{figure}

As  mentioned  earlier,  residual   nuclear  burning  in  white  dwarf
atmospheres can be an important  source of energy, specially for white
dwarfs descending  from progenitors  of low  metallicity.  This  is so
because,   in  the   absence  of   third  dredge-up,   low-metallicity
progenitors   depart    from   the   AGB   with    thicker   envelopes
\citep{Iben_1986},  thus  resulting  in   white  dwarfs  with  thicker
hydrogen  envelopes.   Consequently,   residual  hydrogen  burning  is
expected  to  become relevant  in  white  dwarfs with  low-metallicity
progenitors.  In  particular, \cite{Renedo_2010}  found that  in white
dwarfs resulting from progenitors with sub-solar metallicity, residual
hydrogen burning may account about 30\% of the luminosity when cooling
has     proceeded     down     to    luminosities     ranging     from
$L\,\sim\,10^{-2}\,L_{\sun}$ to $10^{-3}\,L_{\sun}$. This contribution
is  even   more  relevant   when  very  low   metallicity  progenitors
($Z=0.0001$) and  white dwarf masses smaller  than $\sim0.6\,M_{\sun}$
are  considered.   In  these  cases nuclear  reactions  are  the  main
contributor  to the  stellar  luminosity for  luminosities  as low  as
$\log(L/L_{\sun})\simeq-3.2$ \citep{Miller_etal_2013}.   For instance,
consider   an    otherwise   typical   white   dwarf    with   $M_{\rm
WD}=0.6\,M_{\sun}$, descending from a  progenitor with the metallicity
of NGC~6397.  For  this white dwarf it takes about  $1.04$~Gyr to cool
down  to  $\log(L/L_{\sun})\simeq-3.2$  when  no  nuclear  burning  is
considered,  while when  residual nuclear  burning is  considered, the
cooling  age is  $\sim 1.49$~Gyr.   Although for  fainter luminosities
these  differences become  smaller,  residual  hydrogen burning  might
still play  a significant role  in shaping the white  dwarf luminosity
function at  moderately low luminosities.  Given  the distance modulus
of NGC~6397  these luminosities  correspond to apparent  magnitudes of
about $m_{\rm F814W}\approx24\sim26$,  in the hot branch  of the white
dwarf luminosity function.

To assess  the possible effect of  residual nuclear burning we  used a
second  set of  cooling  sequences  where this  source  of energy  was
disregarded. As  before, we  built a grid  of sequences  for different
masses  and  ages and  interpolated  for  the precise  metallicity  of
NGC~6397, keeping constant  the duration of the initial  burst of star
formation,  for  which we  adopted,  as  previously discussed  $\Delta
t=1.0$~Gyr. Afterwards, we ran our  Monte Carlo simulator for the full
range of  the free parameters  and obtained the combined  $\chi^2$ for
each    simulation,    employing    the   procedure    explained    in
Sect.~\ref{subsec:onepop}.  The  white dwarf luminosity  function, the
color-magnitude diagram  and the  color distribution for  the best-fit
model  is  shown  in   Fig.~\ref{f:WDLF_SIN}.   Although  the  general
agreement between the simulated and observed samples for the different
distributions of Fig.~\ref{f:WDLF_SIN} is again quite good, the fit of
some features  of the  observed distributions  is worse  when compared
with  that  obtained  when  residual nuclear  burning  is  taken  into
account.  For instance,  as can be seen  in Figs.~\ref{f:WDLF_CON} and
\ref{f:WDLF_SIN} the small, secondary peak of the observed white dwarf
luminosity function at $m_{\rm F814W}\approx26.75$ is best fitted when
the    sequences    including    residual   hydrogen    burning    are
considered. Moreover, when no residual nuclar burning is considered an
excess    of   simulated    white   dwarfs    redder   than    $m_{\rm
F606W-F814W}\approx1.2$  is  quite apparent  in  the  bottom panel  of
Fig.~\ref{f:WDLF_SIN}. According  to these findings, for  the rest the
calculations presented in  this work we will  only employ evolutionary
sequences  in   which  residual   hydrogen  burning  in   white  dwarf
atmospheres is considered.

\subsection{The initial mass function}
\label{subsec:imf}

\begin{figure}[t]
   \resizebox{1.1\hsize}{!}
   {\includegraphics[width=\columnwidth]{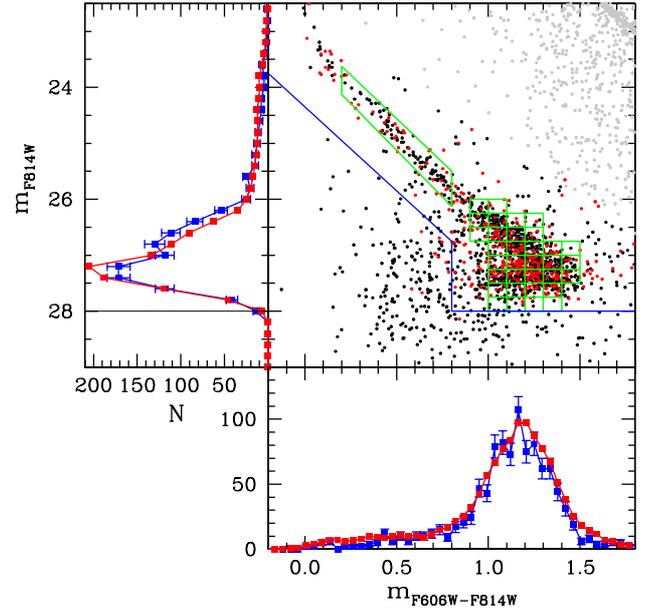}}
   \caption{Same as  Fig.~\ref{f:WDLF_CON} but for the  model in which
     the slope  of the initial mass  function is allowed to  vary (see
     Sect.~\ref{subsec:imf} for details).}
\label{f:WDLF_IMF}
\end{figure}

The initial  mass function is a  key ingredient to model  the observed
properties of globular clusters.  The  most commonly used initial mass
function consists  of a  power law with  a characteristic  power index
$\alpha$, $\Phi(M)\propto M^{-\alpha}$,  where $M$ is the  mass at the
zero age  main sequence.  It is  generally accepted that the  value of
the  power  law  index  should  not  differ  much  from  $\alpha=2.35$
\citep{Salpeter55}. However,  it has been recently  suggested that the
initial mass  function of NGC~6397 could  have a flatter slope  -- see
\citep{Richer2008} and references therein.  In this section we explore
such a possibility.

As   we  did   previously   we  consider   the   model  described   in
Sect.~\ref{subsec:onepop} but now,  we consider $\alpha$ to  be a free
parameter of the fit,  we allow it to vary between  1.35 and 3.35, and
as  we did  before  we  compute the  corresponding  combined value  of
$\chi^2$.   In  Fig.~\ref{f:WDLF_IMF}  we  show  the  result  of  this
procedure.  The best-fit model corresponds to $\alpha=1.95$.  The rest
of  the values  of the  free parameters  for this  best-fit model  are
similar to  that obtained when  a standard Salpeter-like  initial mass
function     with     $\alpha=2.35$      is     adopted     --     see
Sect.~\ref{subsec:onepop}.  Nevertheless, including  $\alpha$ as a new
free parameter does not substantially  improve the fit.  Therefore, we
conclude that,  unless a  more restrictive  observational data  set is
employed, a standard value of $\alpha$ describes adequately the global
observed properties of the white dwarf population of NGC~6397.

\subsection{The fraction of non-DA white dwarfs}
\label{subsec:DBs}

\begin{figure}[t]
   \resizebox{1.1\hsize}{!}
   {\includegraphics[width=\columnwidth]{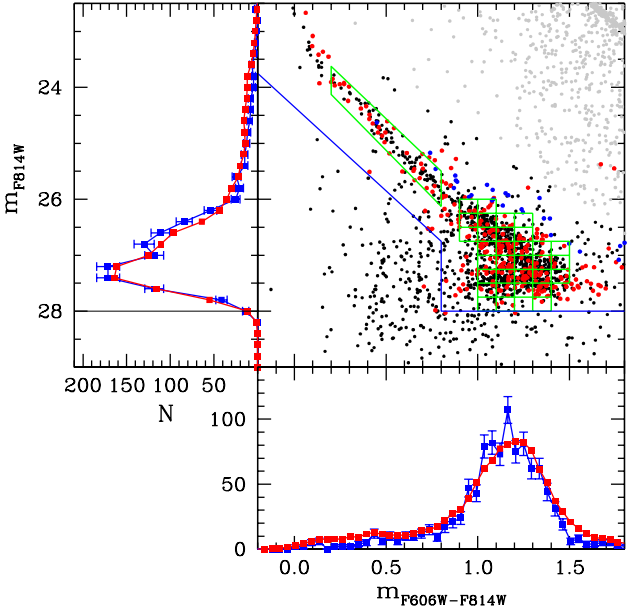}}
   \caption{Same as Fig.~\ref{f:WDLF_CON} but for  a model in which we
     adopt  a   fraction  of   non-DA  white   dwarfs  of   20\%  (see
     Sect.~\ref{subsec:DBs} for details).}
\label{f:WDLF_DBs}
\end{figure}

Although  observations  indicate that  the  fraction  of non-DA  white
dwarfs in NGC~6397 is small, of  the order of $f_{\rm non-DA}\sim 4\%$
\citep{Hansen_2007,Strickler_2009},  it is  worth exploring  which are
its effects on  the color-magnitude diagram, and  on the corresponding
white dwarf luminosity function and  color distribution.  To this end,
in  Fig.~\ref{f:WDLF_DBs}  we show  the  results  when an  unrealistic
fraction  of  20\%  of  non-DA  white  dwarfs  is  adopted.   In  this
color-magnitude  diagram  non-DA  white   dwarfs  are  shown  as  blue
dots. When fractions smaller than 20\%  are used very few non-DA white
dwarfs are found.  Thus, this figure should be regarded  as an extreme
case to illustrate the effects of varying the fraction of non-DA white
dwarfs.

As previously  done in  the preceding sections  we consider  the model
described  in Sect.~\ref{subsec:onepop}  but we  now consider  $f_{\rm
non-DA}$  to be  a  free parameter  of  the fit,  and  we compute  the
corresponding  combined  value  of  $\chi^2$, keeping  fixed  now  the
fraction of unresolved  binaries to 0.04, the best  fit value obtained
previously.  The  value of  the fraction of  non-DA white  dwarfs that
best  fits  the  observed  data is  $f_{\rm  non-DA}=0$.   Hence,  our
analysis  concurs with  observations  that NGC~6397  has a  negligible
fraction of non-DA white dwarfs. However, we note that small fractions
of non-DA white dwarfs are allowed by the present set of observed data
at the 95\% confidence level.

\subsection{Mass segregation in NGC~6397}
\label{subsec:segregation}

\begin{figure}[t]
   \resizebox{1.1\hsize}{!}
   {\includegraphics[width=\columnwidth]{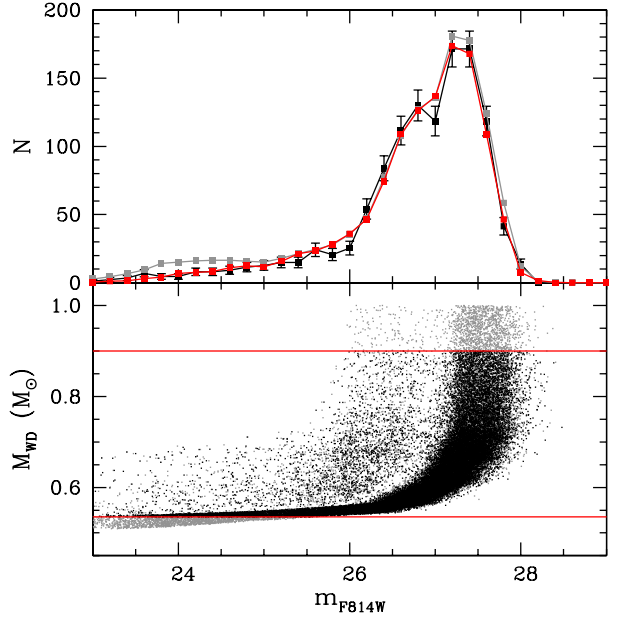}}
   \caption{Top  panel:  Clean  white  dwarf  luminosity  function  of
     NGC~6397  (black line)  compared  with  the synthetic  luminosity
     function for the  entire sample of white dwarfs  (grey line), and
     that obtained after eliminating massive and low-mass white dwarfs
     (red line).  Bottom panel: mass distribution as a function of the
     magnitude for  our simulated sample (black  dots). Also displayed
     in this panel, as horizontal red lines, are the mass cuts.}
\label{f:WDLF_MAS}
\end{figure}

Another important issue worth investigating is the dynamical evolution
of NGC~6397. The dynamical evolution of globular clusters is primarily
governed  by the  gravitational attraction  between individual  stars,
although the interplay of the  evolution of their individual stars and
binary  encounters also  plays an  important role.   The high  stellar
densities in the  cores of globular clusters  favor binary encounters.
In  these  dynamical  interactions   the  more  massive  star  usually
transfers  kinetic  energy  to   the  less  massive  one.  Eventually,
high-mass stars clump in the core  of the cluster, while low-mass ones
move to the outskirts of it, or even escape the cluster.

Given that the  observed sample of white dwarfs employed  in our study
was selected from the ACS field, which is located 5$^\prime$ SE of the
cluster center, it  is reasonable to expect that  mass segregation may
play a role.   Also, the evaporation time of NGC~6397  is $\sim 3$~Gyr
\citep{Heyl2012},  implying that  an important  fraction of  stars, in
particular those with low masses, could have escaped from the cluster.
However,  the  existence of  mass  segregation  in NGC~6397  is  still
controversial   --   see,   for   instance,  the   recent   paper   of
\cite{Martinazzi2014}, and references therein -- and a study employing
white  dwarf stars  would help  in  settling this  issue. Although  an
exhaustive asssessment  of the  dynamical evolution of  NGC~6397 using
the properties  of the white dwarf  population is beyond the  scope of
the  present paper,  a  simple  approach is  worth  pursuing, since  a
comparison  of the  properties of  the simulated  population with  the
observed  sample  could shed  some  light  on  the existence  of  mass
segregation.

To test  how mass  segregation affects the  white dwarf  population of
NGC~6397   we    proceed   as   follows.    The    bottom   panel   of
Fig.~\ref{f:WDLF_MAS} displays the distribution  of white dwarf masses
as a  function of  the magnitude for  our synthetic  population. Black
points  denote typical  carbon-oxygen white  dwarfs, whereas  the grey
points  represent white  dwarfs  with masses  larger  than a  variable
threshold  $M_{\rm u}$  (upper  red  line) and  smaller  than a  given
adjustable  value $M_{\rm  l}$ (lower  red  line).  We  also show  the
observed white  dwarf luminosity function,  which is shown as  a black
line in the top panel of Fig.~\ref{f:WDLF_MAS}, while the grey line is
used  to represent  the luminosity  function of  the entire  synthetic
sample. Finally,  the white dwarf  luminosity function for  stars with
masses $M_{\rm l}<M<M_{\rm  u }$ is represented using a  red line.  As
can  be seen  in  Fig.~\ref{f:WDLF_MAS}, the  bright portion  ($m_{\rm
F814W}\simeq23\sim  26$) of  the luminosity  function is  dominated by
low-mass white  dwarfs. By varying  the lower  mass cut we  fitted the
shape of the hot branch of  the luminosity function, and obtained that
the value  of $M_{\rm l}$  that best  fits the luminosity  function is
$0.535\, M_{\sun}$. On  the other hand, massive  white dwarfs dominate
the  region  beyond   the  maximum  of  the   white  dwarf  luminosity
function. By adjusting the value of the upper cut in masses the height
of the maximum  of the white dwarf luminosity function  can be fitted.
The  upper mass  limit obtained  in  this way  is $0.905\,  M_{\sun}$.
These  two   values  are  those  shown   in  Fig.~\ref{f:WDLF_MAS}  as
horizontal red lines.   Note that the white  dwarf luminosity function
obtained after applying  these two mass cuts (the red  line in the top
panel  of  Fig.~\ref{f:WDLF_MAS})  fits extremely  well  the  observed
luminosity function.   In fact, for  the particular model  analyzed in
Fig.~\ref{f:WDLF_MAS},  which   corresponds  to   a  cluster   age  of
$12.8\,$Gyr, a burst  duration of $1.0\,$Gyr and a  binary fraction of
$10\%$, the value  of the reduced $\chi^2$ is 1.20  when the mass cuts
are applied.  This  number has to be compared with  that obtained when
no mass cuts  are employed, which is $\chi^2=1.67$,  indicating a much
better fit. Obviously,  these cuts reduce size of  the synthetic white
dwarf sample.  For the best-fit mass  cuts this reduction amounts to a
$10\%$.  Hence, we conclude that mass  segregation plays a role in the
fine tunning of the luminosity function.

\section{Conclusions}
\label{sec:con}

In this paper  we studied the white dwarf population  of NGC~6397.  We
have  done  that  employing   an  up-to-date  Monte  Carlo  population
synthesis code,  which incorporates the most  advanced descriptions of
the    observational   biases    and    selection   procedures,    and
state-of-the-art cooling sequences, for the appropriate metallicity of
this cluster.   We have compared  the results of our  simulations with
the most recent observational sample  of the white dwarf population of
this  cluster,  which  was  obtained  using  deep  {\sl  Hubble  Space
Telescope} observations.  In particular, our simulations were compared
with the general  appearance of the color-magnitude  diagram, with the
color distribution and  with the white dwarf luminosity  function in a
quantitative manner, obtaining probability distributions for each pair
of the most important free parameters of the fit. These are the age of
the cluster, the duration of the  initial burst of star formation, and
the fraction of  unresolved binaries.  Also, other  key ingredients in
modelling the white dwarf population  of NGC~6397 were studied.  These
include the adopted reddening and distance of the cluster, the role of
the completeness  of the observed  sample, the fraction  of unresolved
binaries, the  initial-to-final mass  relationship, the  adopted white
dwarf cooling  sequences, the initial  mass function, the  fraction of
non-DA white dwarfs,  and the impact of mass  segregation. Our results
are in  very good agreement  with the  observed data, and  are largely
independent   of   model   assumptions.   However,   the   probability
distributions show  broad, flat maxima that  prevent to simultaneously
obtain accurate determinations  of the age of the cluster,  and of the
duration of its  initial burst of star formation. The  origin of these
flat maxima  is the still  sizable size  of photometric errors  of the
current observed data set.

\begin{table}[t]
\begin{center}
\caption{Age determinations for NGC~6397.}
\label{t:agedet}
\tiny
\begin{tabular}{ll}
\hline
\hline
 $T_{\rm c}$~(Gyr)   & Reference \\
\hline
$12.0\pm0.8$        &  \cite{Twarog2000}    \\
$13.4\pm0.8$        &  \cite{Chaboyer_2001}   \\
$13.9\pm1.1$        &  \cite{Gratton_2003}   \\
$11.47\pm0.47$      &  \cite{Hansen_2007}   \\
$12.0^{+0.5}_{-1.0}$  &  \cite{Winget_2009}   \\
$13.50\pm0.50$      &  \cite{Dotter_2010}   \\
$13.00\pm0.25$      &  \cite{VandenBerg_2013}   \\
$12.8^{+0.5}_{-0.75}$ &  This work   \\
\hline
\end{tabular}
\end{center}
\end{table}

The age of NGC~6397 derived using  the white dwarf cooling sequence is
$12.8^{+0.50}_{-0.75}$~Gyr, when the duration  of the initial burst of
star formation is 1.0~Gyr.  The uncertainty  in the age of the cluster
introduced  by  the  uncertainty  in  the  distance  modulus  is  $\pm
0.3$~Gyr, while  that introduced by  the reddening is  negligible when
compared with  the previous one.   This age estimate agrees  well with
the age  determinations obtained using  the main sequence  turn-off --
see  Table~\ref{t:agedet}.   Specifically,  our age  determination  is
slightly shorter  than the  most recent determinations  obtained using
main  sequence  stars. Specifically,  \cite{VandenBerg_2013}  obtained
$13.00\pm0.25$~Gyr,    and   \cite{Dotter_2010}    derived   $13.50\pm
0.50$~Gyr,  somewhat older  than  previous  estimates.  For  instance,
\cite{Gratton_2003} obtained $13.9\pm1.1$~Gyr and \cite{Chaboyer_2001}
derived $13.4\pm0.8$~Gyr.  However,  our age estimate is  in line with
the age  derived by \cite{Twarog2000}, who  obtained $12.0\pm0.8$~Gyr.
We also find that our age estimate for NGC~6397 is significantly older
than    those   of    \cite{Hansen_2007},   $11.47\pm0.47$~Gyr,    and
\cite{Winget_2009},  $12.0^{+0.5}_{-1.0}$~Gyr,   which  were  obtained
using  the position  of  the  cut-off of  the  white dwarf  luminosity
function.  This  discrepancy can be  partially attributed to  the fact
that we used updated evolutionary  tracks.  This includes not only the
white dwarf cooling sequences, but also the main sequence lifetimes of
white  dwarf progenitors,  and  the  initial-final mass  relationship,
which in our calculations are computed for the appropriate metallicity
of NGC~6397.  However, the use  of reliable evolutionary tracks is not
the only reason  for this, and we stress that  other factors, like the
way in  which the synthetic  population of white dwarfs  is generated,
taking into account  all the known observational  biases and selection
procedures, are also important to obtain reliable ages.

Another interesting  result of  our study  is that  we found  that the
fraction  of  binaries   is  close  to  zero.   This  is  particularly
reassuring, as the percentage of binaries in our best-fit model ($\sim
4.0\%$) is  very similar to that  obtained for main sequence  stars --
see tables~2 and 3 in \cite{Milone2012}.

Having estimated the age and  the percentage of unresolved binaries of
NGC~6397 we decided to study  other important properties of the model.
To start with, we studied the role of hydrogen residual burning in the
atmospheres of white dwarfs with  very metal poor progenitors. To this
end,  we computed  a complete  set of  full evolutionary  sequences in
which residual nuclear burning was (artificially) not considered.  The
results of this numerical experiment  indicate that although good fits
to the observed  properties of the white dwarf  population of NGC~6397
can also be obtained using this set of evolutionary tracks, the fit is
better when the  cooling sequences in which  residual hydrogen burning
is considered.   This result  sheds light on  the importance  of third
dredge-up and  extra mixing episodes in  low-mass, low-metallicity AGB
stars,  for  which  theoretical  and  observational  evidence  is  not
conclusive about  their occurrence.  As shown  in \cite{camissasa}, in
the absence of third dredge-up episodes  during the AGB phase, most of
the evolution  of white  dwarfs resulting from  low-mass ($M  < 1.25\,
M_{\sun}$),  low-metallicity   progenitors  is  dominated   by  stable
hydrogen burning.  In view of our result that low-mass white dwarfs in
NGC~6397  are  expected  to   sustain  significant  residual  hydrogen
burning, we  conclude that their  low-mass progenitors would  not have
suffered  from the  carbon-enrichment  due to  third dredge-up  during
their AGB evolution.

Another  possible concern  could  be  the slope  of  the initial  mass
function.  To assess  this point we conducted a  series of simulations
in which we  varied the power-law index of the  initial mass function,
and we  found that that although  a shallower slope will  fit slightly
better the  properties of the  observed sample, the standard  value of
the Salpeter mass function fits equally well such properties, within a
95\%  confidence  level. Also,  we  studied  which  is the  impact  of
adopting a  different fraction of  non-DA white dwarfs and  found that
our best  fit model  corresponds with a  negligible fraction  of these
stars, in agreement with observations.  Finally, we made a preliminary
test to  assess the  effects of  mass segregation  on the  white dwarf
population of NGC~6397.   By adjusting the bright and  the dim portion
of the  luminosity function we  obtained that white dwarf  with masses
larger than  $0.905\, M_{\sun}$ or smaller  than $0.535\,M_{\sun}$ are
absent  in the  observed sample,  possibly  as a  consequence of  mass
segregation.  When  this  is  taken  into  account  in  the  simulated
populations  the agreement  between the  observed and  the theoretical
data turns out to be excellent.

In  summary,  we  have  conducted the  most  complete  and  up-to-date
population synthesis study  of the white dwarf population  of the old,
metal-poor, globular  cluster NGC~6397. This  study has allowed  us to
derive useful constraints on the  characteristics of this cluster, and
to obtain  an estimate of its  age. Nevertheless, we emphasize  that a
better  observational data  set, with  smaller photometric  errors for
white dwarfs of  magnitudes $\ga 27$ will help in  obtaining even more
reliable estimates of both the age of its star formation history.


\begin{acknowledgements}
This    research   was    partially   supported    by   MCINN    grant
AYA2014--59084--P, by  the European  Union FEDER  funds, by  the AGAUR
(Spain),  by  the  AGENCIA  through the  Programa  de  Modernizaci\'on
Tecnol\'ogica  BID 1728/OCAR,  and by  the PIP  112-200801-00940 grant
from CONICET (Argentina). We also thank B.M.S. Hansen, for stimulating
discussions and for providing us with the observed data.
\end{acknowledgements}

\bibliographystyle{aa}
\bibliography{6397}

\end{document}